\documentclass[12pt]{article}
\usepackage[dvips]{color}
\usepackage{amsmath}
\usepackage{graphicx}
\usepackage{subfigure}
\usepackage{epsfig}
\usepackage{amsmath}
\usepackage{cite}
\usepackage{color}
\usepackage{subfigure}

\begin{document}
\begin{center}
\Large{\bf WGC and WCC  for charged black holes with quintessence and cloud of strings}\\
\small \vspace{1cm} {\bf Mohammad Reza Alipour $^{\star}$\footnote {Email:~~~mr.alipour@stu.umz.ac.ir}}, \quad
{\bf Jafar Sadeghi$^{\star}$\footnote {Email:~~~pouriya@ipm.ir}} and \quad
{\bf Mehdi Shokri $^{\dagger}$\footnote {Email:~~~m.shokri@umz.ac.ir}}\\
\vspace{0.5cm}$^{\star}${Department of Physics, Faculty of Basic
Sciences,\\
University of Mazandaran
P. O. Box 47416-95447, Babolsar, Iran}\\
\vspace{0.5cm}$^{\dagger}${School of Physics, Damghan University, P. O. Box 3671641167, Damghan, Iran}\\
\end{center}
\begin{abstract}
One of the important problems with the existence of   weak gravity conjecture ($WGC$) is the violation of  the cosmic censorship. Such a cosmic phenomena is important and consistent in general relativity. It means that  for a charged black hole in four dimensions and in the normal state, the $WGC$ cannot hold due to the violation of the weak cosmic censorship conjecture ($WCCC$).
But in this paper, in order to eliminate such a violation, we try to consider RN black hole at two cases, with quintessence,  quintessence and cloud of strings in asymptotically flat and non-asymptotically flat. By using such metric background and applying restrictions on its parameters, we show that there will be a good compatibility between $WGC$ and $WCCC.$ Here, we can also see that quintessence case has less freedom in establishing compatibility than quintessence and cloud of strings. Because in the second case, the number of parameters is more and there is a degree of freedom to better establish  the compatibility of two theories. Also, here we explain  charged black hole in \emph{presence of quintessence and cloud of string} for $Q>M$ and  use  some interesting figures we show the corresponding  the black hole has an event horizon.
\end{abstract}
\section{Introduction}\label{s1}
As we know, recently  several researchers with many efforts investigate  quantum gravity from different approaches.
One of these approaches  is  swampland program, which is used to determine the boundary between low-energy effective field theories compatible with quantum gravity (landscape) and low-energy effective field theories that are not compatible with quantum gravity (swampland)\cite{1,2,3,4}. Here we note that, one of the main conditions in swampland's program is that there is no global symmetry in quantum gravity, but gauge symmetry can exist\cite{1,2}. Generally, one can say that the above mentioned condition leads us to  weak gravity conjecture. By using such conjecture, we can find the effective field theories compatible with quantum gravity. Here, we note that in order to achieve the weak gravity conjecture, the ratio of electric charge to mass must be  greater than one $\left(\frac{q}{m}>1\right)$. So, in this case , gravity is considered as weakest force \cite{5,6}.
Of course, there are other conjectures in the swampland program and it  helps  to find effective fields theories compatible with quantum gravity, for further study refer to \cite{1,2,6,7,8,9,10}.
On the other hand, the weak cosmic censoring conjecture is very important for theoretical physics. Because, there is an unavoidable fundamental singularity inside the black hole, near this singularity, the laws of physics fail. To avoid this phenomenon, Penrose proposed the weak cosmic censorship conjecture \cite{10a}. This conjecture claims, the singularity in black hole space-time should be hidden by the event horizon.
We cannot use the weak gravity conjecture for the Reissner-Nordstr\"{o}m ($RN$) black hole because when $Q>M$, it violates the cosmic weak censoring conjecture, and therefore these two conjectures are incompatible with each other.
But, when the $RN$ black hole decay in the extremality state, according to the energy conservation theorem, there are decay products whose charge-to-mass ratio is greater than one.
In this case, due to the violation of weak cosmic censorship, we cannot consider them as black holes, but we consider products as particles \cite{2,11}.  So, this is one of most important problem for the weak gravity conjecture. In that case also, many articles have been written in these years to prove the weak gravity conjecture \cite{12,13,14,15}.\\
Here, generally one can say that, when it comes to the WGC, one of the relevant questions is: which states are expected to be the weak gravity states? Said differently, do we expect the WGC to be satisfied by particles or other black holes? Of course, black holes of pure Einstein-Maxwell theory cannot satisfy the WGC (have $Q > M$) because they would then have to violate WCCC (requiring
$Q < M$  for charged BHs \footnote{The event horizons of the charged black hole are obtained as $r_\pm=M\pm\sqrt{M^2-Q^2}$. So, when $Q > M$ the black hole will not have an event horizons, in which case it will have a naked singularity which will lead to WCCC violation.}). However, corrections to vanilla Einstein-Maxwell theory can lead to corrections of the BH solutions allowing them to tend towards super-extremality (i.e. have
$Q > M$) without violating WCC. This has been an active area of discussion recently, see for example \cite{15a,15b,15c,15d}.\\
On the other hand, one of the most important discoveries in the field of modern cosmology is the finding the  observational confirmation of the accelerated expansion of the universe \cite{16}. We can also consider this accelerated expansion of the universe as a result of gravitational repulsive energy, which is related to negative pressure on a cosmological scale.
Regarding this negative pressure, there are various answers, such as considering the cosmic constant which is related to the vacuum energy in Einstein's equations. Here also, we note that by  making some corrections in the cosmic geometry one can assume the corresponding  expansion is caused by a fluid called quintessence dark energy  \cite{17,18,19}.
Therefore,  surrounding the black hole by quintessence will  be important in an astronomical scenario \cite{20,21,22,23}.
Also,  when the Schwarzschild black hole is surrounded by a spherically symmetric cloud of strings,  the event horizon of the corresponding  black hole will be changed by such correction \cite{24}.\\
Our motivation in this article to be able to solve one of the problems of  weak gravity conjecture, which is in conflict with the weak cosmic censorship conjecture. In order to solve this problem we consider the \emph{quintessence and cloud of strings} around the charged black hole.\\
The layout of the paper is the following.
In section 2, we introduce the $RN$ black hole metric (or the charged black hole) in the presence of the quintessence and cloud of strings.
In section 3, we consider the $RN$ black hole metric only in the presence of quintessence and analyze the black hole event horizons in different situations. By using diagrams and information of asymptotic and non-asymptotic flat space we  find out under what conditions the $WGC$ is compatible with the $WCCC$.
In section 4,   we examine the $RN$ black hole metric in the presence of quintessence and cloud of strings.  By obtaining black hole event horizons and considering also asymptotic and non-asymptotic flat space,  we investigate the connection between weak gravity conjecture and weak cosmic censorship. Finally, we describe the results in section 5.
\section{Charged  black holes in presence of quintessence and cloud of strings}\label{s2}
Now we are going to introduce the charged black hole with  presence of quintessence and cloud  strings which is given by \cite{25,26},
\begin{equation}\label{eq1}
dS^2=f(r)dt^2-f^{-1}(r)dr^2-r^2(d\theta^2+\sin^2(\theta)d\varphi^2)
\end{equation}
and $f(r)$ is,
\begin{equation}\label{eq2}
f(r)=1-b-\frac{2M}{r}+\frac{Q^2}{r^2}-\frac{c}{r^{3\omega_q+1}}
\end{equation}
where $M$, $Q$, $\omega_q$, $c$ and  $b$ are the mass, electric charge of the black hole, the quintessential state parameter, the
positive normalization factor and  constant which takes care of the presence of the cloud of strings, respectively. Here also, one can obtain the energy density (positive) and isotropic pressure (negative) as follows \cite{27,28},
\begin{equation}\label{eq3}
\rho_q=-\frac{c}{2}\frac{3\omega_q}{r^{3(\omega_q+1)}},  \qquad  p_q=\rho_q \omega_q
\end{equation}
As we know $\omega_q$ can be placed in range $-1<\omega_q<0$, so we can  divide it into two ranges as  $-\frac{1}{3}\leq \omega_q <0$ which is asymptotically flat solutions and  $-1<\omega_q<-\frac{1}{3}$ which is non-asymptotically flat solutions \cite{22,26}. Also here, we note that  in case of  $\omega_q <-1$ we have   phantom dark energy and in case of $-1 < \omega_q<-\frac{1}{3}$ we have the quintessence dark energy.
In the following, we examine the black hole in two ranges, asymptotic flat and  non-asymptotic flat space.
\section{Investigation of charged black hole  with quintessence}\label{s3}
In this section, we investigate the charged black hole metric only in the presence of quintessence in two ranges of asymptotic and non-asymptotic flat space. By setting $f(r)=0, b=0$ and $v=\frac{1}{r}$, we can obtain the event horizon of the black hole as,
\begin{equation}\label{eq4}
1-2M v+Q^2 v^2=c v^{3\omega_q+1}
\end{equation}
To  solving equation \eqref{eq4},  we have some problem, because such equation is  complicated. In order to solve such equation we employ some figures and discuss different cases. First of all we consider
 $-\frac{1}{3}\leq \omega_q <0$, as a asymptotically flat space.
\begin{figure}[hbt!]
\includegraphics[width=.28\textwidth]{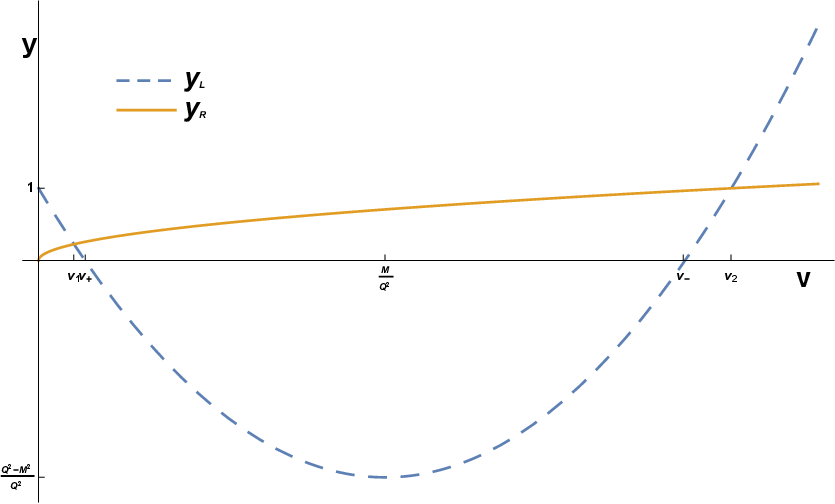}
(a)
\includegraphics[width=.28\textwidth]{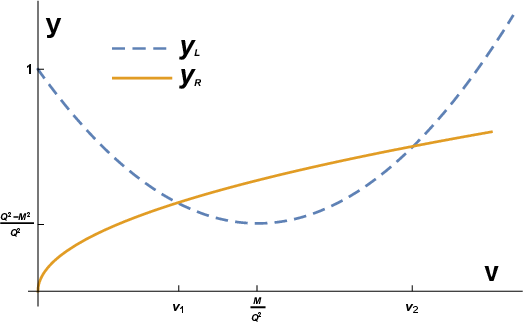}
(b)
\includegraphics[width=.28\textwidth]{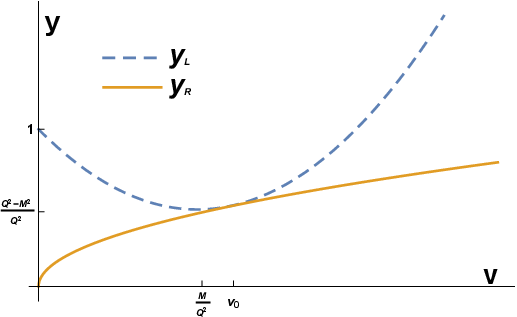}
(c)
\caption{Plots of $y_{L} = 1-2M v+Q^2 v^2$  and $y_{R} = c v^{3\omega+1}$ for $-1/3\leq \omega_q < 0$.} \label{fig:8a}
\end{figure}
According to equation \eqref{eq4}, we draw two graphs $y_{L} = 1-2M v+Q^2 v^2$  and $y_{R} = c v^{3\omega+1}$ as figure 1, for $\frac{Q^2}{M^2}\leq 1$ and  $\frac{Q^2}{M^2}> 1$  cases. In  the figure 1 (a), when $y_R=0$, the $y_L$ graph intersects the $v$ axis at two points $(v_+, v_-)$, which are the  inner and outer event horizons of the ordinary RN black hole. Here, also graph $y_L$ has a minimum value as $\frac{Q^2-M^2}{M^2}$ at the point of $v_{min}=\frac{M}{Q^2}$. Also, we see in figure 1(a), when $\frac{Q^2}{M^2}\leq 1$, two graphs $y_R$ and $y_L$  intersect at two points $(v_1, v_2)$. So,  the outer event horizon of the RN black hole in the presence of quintessence $r_1=\frac{1}{v_1}$ is larger than the outer event horizon of ordinary RN black hole $r_+=\frac{1}{v_+}$ and also the inner horizon becomes smaller. As you can see in figure 1(b), when $\frac{Q^2}{M^2}> 1$, the parabola $y_L$ does not intersect the v-axis,  this means the ordinary RN black hole has a naked singularity, which violates the $WCCC$ and the $WGC$ is not established. But in the presence of $quintessence$, two diagrams $y_R$ and $y_L$  intersect to each other at two points $v_1$ and $v_2$. It means that the black hole has two internal and external event horizons, despite the fact that it is$\frac{Q^2}{M^2}> 1$, the $WCCC$ is not violated, also we have  $WGC$. In this situation, the $WGC$ and the $WCCC$ are compatible with each other without any tension. In figure 1(c), two graphs $y_R$ and $y_L$  are tangent to each other at a common point. In that case, we can define  the extremal state of the black hole at the point of $v_0$, with the $\frac{Q^2}{M^2}> 1$ condition and also  we consider $c=c_e$.
When the parameters $\omega_q, M, \frac{Q^2}{M^2}> 1$ are constant, if  $c<c_e$, then the black hole has a naked singularity and violates the weak cosmic censorship. And  if  $c>c_e$, the black hole has two event horizons, so the $WCCC$ is not violated. The case $c=c_e$, we have an extreme black hole with event horizon $r_0=\frac{1}{v_0}$.
Now, by using equation \eqref{eq4} and  tangent of the two graphs of $y_R$ and $y_L$ we obtain the values of $v_0$ and $c_e$  ,
\begin{equation}\label{eq5}
1-2M v_0+Q^2 v_0^2=c v_0^{3\omega_q+1}
\end{equation}
\begin{equation}\label{eq6}
-2M +2Q^2 v_0=c(3\omega_q+1) v_0^{3\omega_q}
\end{equation}
By placing equation \eqref{eq6} in equation \eqref{eq5}, the values of  $v_0$ and $c_e$  are obtained by following expressions,
\begin{equation}\label{eq7}
v_0=\frac{\sqrt{9\omega_q^2M^2+(1-9\omega_q^2)Q^2}-3\omega_qM}{(1-3\omega_q)Q^2}  \qquad   c_e=\frac{2(Q^2v_0-M)}{(1+3\omega_q)v_0^{3\omega_q}}
\end{equation}
According to the above relationship, conditions $c>0$ and $-1/3<\omega_q<0$, we find $v_0>M/Q^2$, which is shown in figure 1(c).

Also, by fixing ($Q, M,c$) and increasing $\frac{Q^2}{M^2}$, we can reach the extremity black hole, which has such a relationship $\frac{Q^2}{M^2}=\left(\frac{Q^2}{M^2}\right)_e$. By fixing $c \ll 1, \omega_q, M$ and using some approximation for the RN extreme black hole  in the presence of quintessence,  we  have  following relation  \cite{26},
\begin{equation}\label{eq8}
\left(\frac{Q^2}{M^2}\right)_e=1+\frac{c/2}{M^{3\omega_q+1}}+\mathcal{O}\left(\frac{c/2}{M^{3\omega_q+1}}\right)^2 \qquad  (c\ll1)
\end{equation}
According to relations \eqref{eq2} and \eqref{eq4} for the value of $\omega_q=-1/3$, the extremity black hole relation corresponds to,
\begin{equation}\label{eq9}
\left(\frac{Q^2}{M^2}\right)_e=\frac{1}{1-c} \qquad for \qquad c \ll 1 \qquad \Rightarrow  \qquad \left(\frac{Q^2}{M^2}\right)_e=1+c+\mathcal{O}(c^2)
\end{equation}
The relations \eqref{eq8} and \eqref{eq9}, lead us to find that  $WGC$ and $WCCC$ are established for the RN black hole in the \emph{presence of quintessence} as well as its \emph{extremality state}.\\
In second case, we consider $-1<\omega_q<-\frac{1}{3}$ as a non-asymptotically flat space.
Also here, to solve the equation \eqref{eq4} and find the event horizons of the black hole, in the range of $-1<\omega_q<-\frac{1}{3}$, we use graphical and analytical methods.
\begin{figure}[hbt!]
\includegraphics[width=.45\textwidth]{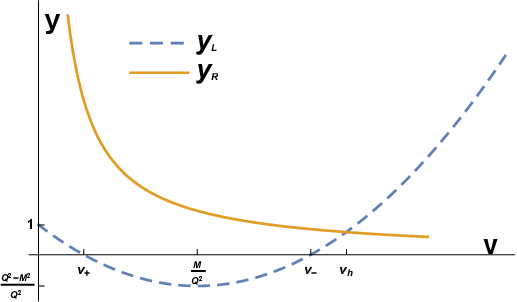}
(a)
\includegraphics[width=.45\textwidth]{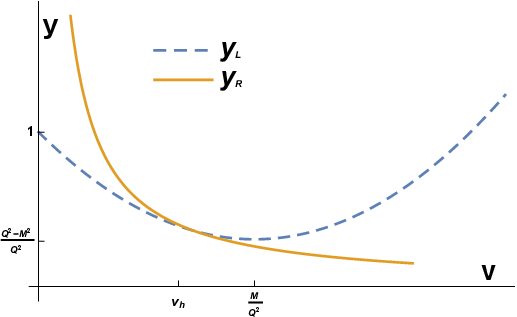}
(b)
\caption{Plots of $y_{L} = 1-2M v+Q^2 v^2$  and $y_{R} = c v^{3\omega+1}$ for $-1<\omega_q<-\frac{1}{3}$.} \label{fig:8a}
\end{figure}
In figure 2(a), we  can see that when $\frac{Q^2}{M^2}\leq 1$,  two graphs of $y_{L}$ and $y_{R}$ intersect at the $v_h$ point. Therefore, we find that the $RN$ black hole has one event horizon in the presence of quintessence, and its value is smaller than the internal event horizon of an ordinary $RN$ black hole $r_h<r_-=M-\sqrt{M^2-Q^2}$. For $\frac{Q^2}{M^2}\geq 1$ in figure 2(b),  two graphs intersect at a point $v_h$ whose value is smaller than $\frac{M}{Q^2}$ ($v_h < \frac{M}{Q^2}$). We find that the black hole $RN$  in the \emph{presence of quintessence} at range of $-1<\omega_q<-\frac{1}{3}$, when $\frac{Q^2}{M^2}\geq 1$  has one event horizon, unlike the ordinary $RN$ black hole, it does not violate the $WCCC$.  Also here we note that the $WGC$  is satisfied by the corresponding black and above mentioned range.\\
In the case $c\ll 1$, for both the $\frac{Q^2}{M^2}\geq 1$ and $\frac{Q^2}{M^2}\leq 1$, two graphs intersect at three points($v_1, v_2, v_3$), where the black hole has three event horizons, and the outer event horizon $v_1$ tends to infinity, as shown in figure 3.
\begin{figure}[hbt!]
\includegraphics[width=.45\textwidth]{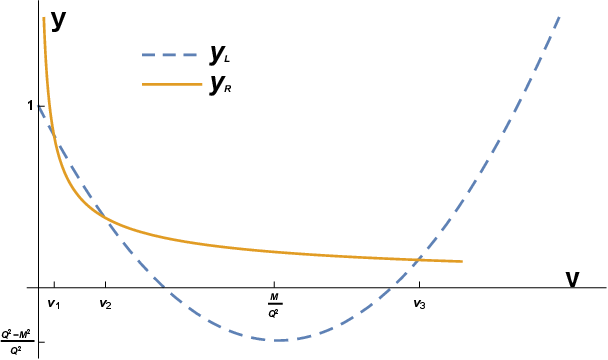}
(a)
\includegraphics[width=.45\textwidth]{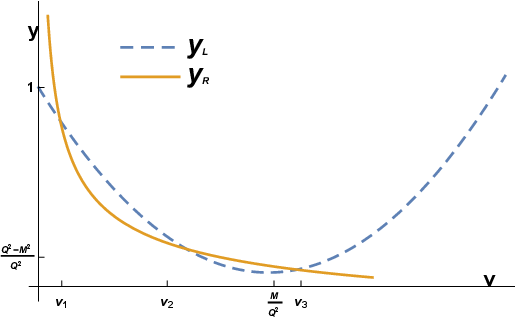}
(b)
\caption{Plots of $y_{L} = 1-2M v+Q^2 v^2$  and $y_{R} = c v^{3\omega+1}$ for $-1<\omega_q<-\frac{1}{3}$ and $c\ll 1$.} \label{fig:8a}
\end{figure}
To find the extremality state of the black hole, we use the conditions of $T=\frac{\partial f/\partial r}{4\pi}$ and equation \eqref{eq4}. It leads to the same equations \eqref{eq5} and \eqref{eq6} in the previous section, so the answer is similar to equation \eqref{eq7}, with the difference that, for $c>0$ and $-1<\omega_q<-\frac{1}{3}$, we have the condition $v_0<\frac{M}{Q^2}$. According to relation \eqref{eq7}, if we take $\frac{Q^2}{M^2} > 1$ , we have an extremal black hole. By fixing ($M, Q, \omega_q$) and changing $c$ near to the $c_e$, we examined equation \eqref{eq4} and black hole event horizons in figure 4.
\begin{figure}[hbt!]
\includegraphics[width=.7\textwidth]{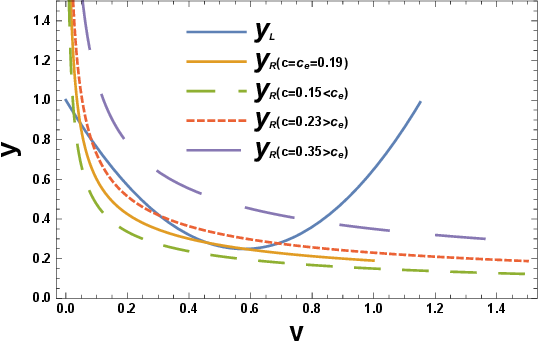}
\caption{Plots of $y_{L} = 1-2M v+Q^2 v^2$  and $y_{R} = c v^{3\omega+1}$ for $\omega_q=-0.5$, $M=1.3$ and $Q=1.5$.} \label{fig:8a}
\end{figure}
For $c < c_e$, two graphs $y_{L}$ and $y_{R}$ intersect at one point, which means the black hole has an event horizon, which increases as $c$ decreases, but when  $c>c_e$, we have two situations. In the first case, if $c$ is slightly more than $c_e$, the two graphs intersect at three points. And in the second case, when $c\gg c_e$, two graphs intersect at only one point, which is the value of the event horizon of the black hole $r_h<\frac{Q^2}{M}$. Also, with the increase of $c$, the radius of the event horizon becomes smaller. By fixing ($\omega_q, M, c$) and increasing $\frac{Q^2}{M^2}$, we can use relation \eqref{eq7} to obtain the extremal black hole in the presence of quintessence for $c\ll 1$,
\begin{equation}\label{eq10}
\left(\frac{Q^2}{M^2}\right)_e=1+\frac{c}{2}M^{-(3\omega_q+1)}+\mathcal{O}\left(\frac{c}{2}M^{-(3\omega_q+1)}\right)^2 \qquad  (c\ll1)
\end{equation}
Equation \eqref{eq10} is similar to equation \eqref{eq8}, but the range of $\omega_q$ is different.
When $c\rightarrow 0$, the equations \eqref{eq8} and \eqref{eq10}  transformed into the extremal state of the ordinary RN black hole i.e. $\frac{Q^2}{M^2}=1$. This state is not established in equation \eqref{eq8} when $M\ll 1$ and in equation \eqref{eq10}  when $M\gg 1$ which is caused by the $3\omega_q+1$ sign for two ranges of $\omega$.
Using the above diagrams and calculations, it is possible to show the difference between two black holes in the presence of quintessence with asymptotically  flat and non asymptotically flat solutions. Thus, for  asymptotically flat solution in $\frac{Q}{M}>1$ condition, there is an event horizon only if the quintessence parameter has a lower bound, which is obtained by equation (7), in extreme case. Therefore, only when $c\geq c_{e}$, the WGC and WCCC are compatible to each other. But for non-asymptotically flat for the condition $\frac{Q}{M}>1$, we have event horizon for all values of the quintessence parameter $c$. In this case, for all value of $c$, there is compatibility between two theories as WGC and WCCC.
\begin{table}
  \centering
\begin{tabular}{|c|c|c|c|}
  \hline
   & $c>c_e$ & $c=c_e$ & $c<c_e$ \\
   \hline
  $-\frac{1}{3}\leq \omega_q<0$ & Yes & Yes  & No \\
  \hline
  $-1< \omega_q<-\frac{1}{3}$ & Yes & Yes & Yes \\
  \hline
\end{tabular}
\caption{WGC and WCCC compatibility conditions for the RN black hole in the presence of quintessence (Yes=compatible  and No=incompatible).}\label{20}
\end{table}
\section{Investigation of charged black hole with quintessence and cloud of strings}\label{s4}
In this section, we investigate the black hole  in the presence of quiescence and cloud of strings. We employ such metric background and find   the relationship between the weak gravity conjecture and the weak cosmic censoring conjecture. Here also,  by considering $f(r)$ and $r=\frac{1}{v}$, we will try to examine black hole event horizons for $\frac{Q^2}{M^2}>1$ and $\frac{Q^2}{M^2}<1$ conditions.  In that case, we use some graphs and relationships in two ranges of asymptotic and non-asymptotic flat space. Now, we start following  $f(r)$ for the case $-\frac{1}{3}\leq \omega_q <0$, asymptotically flat, so we have
\begin{equation}\label{eq11}
1-2M v+Q^2 v^2=b+c v^{3\omega_q+1}
\end{equation}
In case of $-\frac{1}{3}\leq \omega_q <0$, asymptotically flat, we have some figures.
\begin{figure}[hbt!]
\includegraphics[width=.28\textwidth]{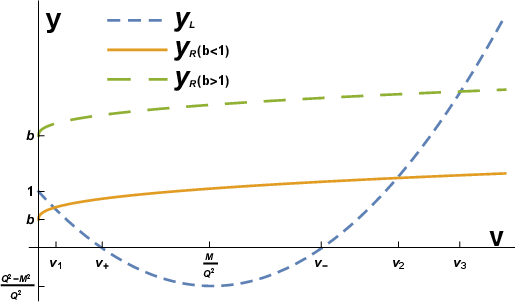}
(a)
\includegraphics[width=.28\textwidth]{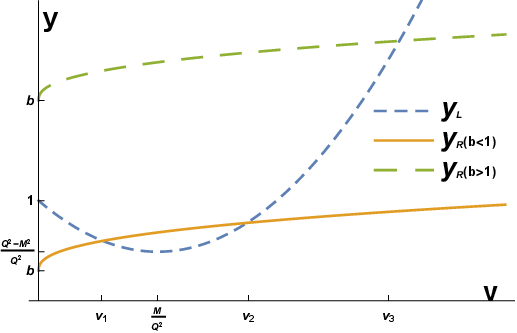}
(b)
\includegraphics[width=.28\textwidth]{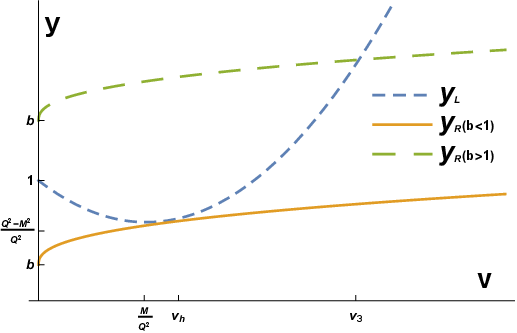}
(c)
\caption{Plots of $y_{L} = 1-2M v+Q^2 v^2$  and $y_{R} = b+c v^{3\omega+1}$ for $-\frac{1}{3}<\omega_q<0$.} \label{fig:8a}
\end{figure}
Figure 5(a) shows that for $\frac{Q^2}{M^2}<1$, when $b<1$, the black hole has two event horizons $(v_1,v_2)$, but for $b>1$, we have only one event horizon $v_3$, which decreases as $b$ increases. We know that the usual  RN black hole violates the $WCCC$ when $\frac{Q^2}{M^2}>1$.
As shown in figure 5(b), when  $\frac{Q^2}{M^2}>1$,  two graphs intersect at two points ($v_1,v_2$) for $b<1$ and at one point $v_3$ for $b>1$, which means that the black hole has two event horizons and one event horizon, respectively. Therefore, we can conclude that \emph{the black hole in the presence of quintessence and cloud of strings satisfies both the conditions of the $WGC$ and the $WCCC$.}
Figure 5(c) shows the extremality state of the black hole. It  can be obtained  by  equation \eqref{eq11} and  slope of the two graphs $y_R$ and $y_L$  as follows,
\begin{equation}\label{eq12}
1-2M v_h+Q^2 v_h^2=b+c v_h^{3\omega_q+1}
\end{equation}
\begin{equation}\label{eq13}
-2M +2Q^2 v_h=c(3\omega_q+1) v_h^{3\omega_q}
\end{equation}
 By using equations \eqref{eq12} and \eqref{eq13}, we obtain the event horizon of the black hole $v_h$ and the $c_e$,
\begin{equation}\label{eq14}
v_h=\frac{\sqrt{9\omega_q^2M^2+(1-9\omega_q^2)(1-b)Q^2}-3\omega_qM}{(1-3\omega_q)Q^2}  \qquad   c_e=\frac{2(Q^2v_h-M)}{(1+3\omega_q)v_h^{3\omega_q}}
\end{equation}
From figure 5(b) and equation \eqref{eq14}, we can see that for a black hole, if $b$ larger than one ($b>1$), we will not have an extremal state. And if $b$ smaller than one ($b<1$), we will have an extremal state.
Also, for $b < 1$ and $Q^2/M^2>1$ when $c < c_e$, the two graphs do not intersect each other. In this case, we will not have an event horizon, so WCCC and WGC are not compatible.
Also here, using equation \eqref{eq14} and figure 5(c), it can be shown that we have an extremal state for black hole only for $b<1$, whose event horizon is $r_h<\frac{Q^2}{M}$. By comparing the equations \eqref{eq14} and \eqref{eq8} to each other, we get that the presence of cloud string near to black hole increases the radius of event horizon in extremal state.
By fixing $c,M,b,\omega_q$ and increasing $Q^2/M^2$, it is possible to reach the extremal state of the black hole. By  using equation \eqref{eq14} for the $b\ll 1$ and $c\ll 1$ conditions one  can obtain following  equation ,
\begin{equation}\label{eq15}
\left(\frac{Q^2}{M^2}\right)_e=1+b+\frac{c/2}{M^{3\omega_q+1}}+\mathcal{O}\left[\left(\frac{c/2}{M^{3\omega_q+1}}\right)^2+\frac{bc/2}{M^{3\omega_q+1}}\right] \qquad  (c\ll 1,b\ll 1)
\end{equation}
 Also using relation \eqref{eq12} for the $\omega_q=-1/3$ condition, the extremal state in the presence of quintessence and cloud of strings will be as,
\begin{equation}\label{eq16}
\left(\frac{Q^2}{M^2}\right)_e=\frac{1}{1-b-c} \qquad for \qquad (c \ll 1, b \ll 1) \qquad \Rightarrow  \qquad \left(\frac{Q^2}{M^2}\right)_e=1+b+c+\mathcal{O}(b+c)^2
\end{equation}
Therefore according relations \eqref{eq15} and \eqref{eq16} ,for \emph{the extremality of the RN black hole in the presence of quintessence and cloud of strings, the weak gravity conjecture and the weak cosmic censorship conjecture are consistent and do not violate each other.}
In the second case one can consider the $-1<\omega_q<-\frac{1}{3}$, non-asymptotically flat and we have some figures.
\begin{figure}[hbt!]
\includegraphics[width=.45\textwidth]{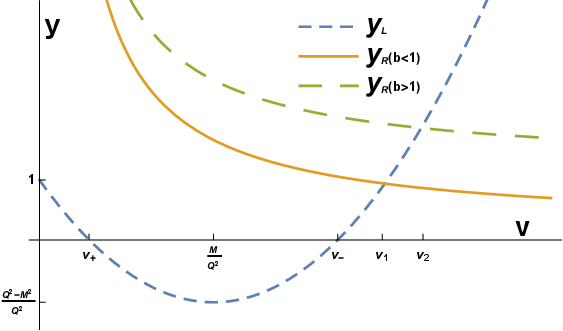}
(a)
\includegraphics[width=.45\textwidth]{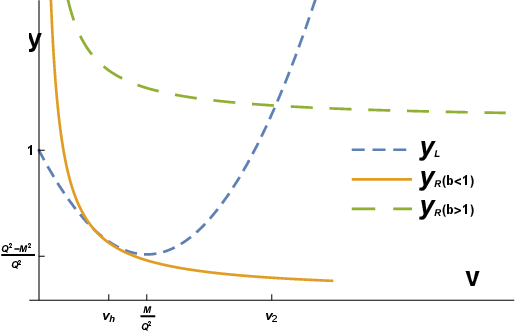}
(b)

\caption{Plots of $y_{L} = 1-2M v+Q^2 v^2$  and $y_{R} = b+c v^{3\omega+1}$ for $-1<\omega_q<-\frac{1}{3}$.} \label{fig:8a}
\end{figure}
As we can see in figure 6, for all values of $b$ in any mode and range $-1<\omega_q<-\frac{1}{3}$, we can have at least one event horizon for the black hole. Also, if we have the $\frac{Q^2}{M^2}>1$ and $c,b<1$ conditions, we can define an extremality state for the black hole, whose relationship is obtained by following equation,
\begin{equation}\label{eq17}
\left(\frac{Q^2}{M^2}\right)_e=1+b+\frac{c}{2}M^{-(3\omega_q+1)}+\mathcal{O}\left[\left(\frac{c}{2}M^{-(3\omega_q+1)}\right)^2+\frac{bc}{2}M^{-(3\omega_q+1)}\right] \qquad  (c\ll1, b\ll 1)
\end{equation}
The above relation is very similar to relation \eqref{eq15}, except that the $3\omega_q+1$ value for the $-1<\omega_q<-\frac{1}{3}$ range is negative.\\
As we have seen, relations \eqref{eq15}, \eqref{eq16} and \eqref{eq17} are extrimality states of black hole in presence of quintessence and cloud of strings. On the other hand, we showed that relations \eqref{eq8}, \eqref{eq9} and \eqref{eq10} are extrimality states of black hole in presence of only quintessence. We find here that in the first case there is more freedom parameter $b$ to establish compatibility between WGC and WCCC.
\begin{table}
  \centering
\begin{tabular}{|c|c|c|c|c|}
  \hline
   & $c>c_e$ & $c=c_e$ & $c<c_e$ &  \\
  \hline
  $-\frac{1}{3}\leq \omega_q<0$ & Yes & Yes & No & $b<1$ \\
  \hline
  $-\frac{1}{3}\leq \omega_q<0$ & Yes & Yes & Yes & $b\geq 1$ \\
  \hline
  $-1< \omega_q<-\frac{1}{3}$ & Yes & Yes & Yes & $b<1$ \\
  \hline
 $-1< \omega_q<-\frac{1}{3}$ & Yes & Yes & Yes & $b\geq 1$ \\
  \hline
\end{tabular}
\caption{ WGC and WCCC compatibility conditions for the RN black hole in the presence of quintessence and cloud of strings (Yes=compatible  and No=incompatible).}\label{22}
\end{table}
\section{Conclusions}\label{s5}
As we know, one of the problems of WGC is its incompatibility with WCCC in usual black hole . In this article, we have shown that when we put the black the presence of quintessence and cloud of strings, both in asymptotically flat and non asymptotically flat state, there is compatibility between WGC and WCCC.  In that case,  we found here that when the charge - to - mass ratio is greater than one($Q>M$) we have an event horizon. But for usual RN black hole with that ratio, we do not have an event horizon, so  two conjectures are not  compatible with each other.
Even, in the extremality state for these black holes, we can have consistency in the weak gravity conjecture for $b$  much smaller than one ($b\ll 1$) and $c$ much smaller than one ($c\ll 1$).
Here, we note that different between asymptotically flat and non-asymptotically flat for the above mentioned black holes will be as: In extremality state of black hole for the case of asymptotically flat we have $-\frac{1}{3}\leq \omega_q <0$ and $M\ll 1$. Also, for the non-asymptotically flat we have $-1\leq \omega_q <-\frac{1}{3}$ and $M\gg 1$, so the probability of WGC condition is very low.
Also here we note that, it seems that black holes can not be good candidates for dark energy and dark matter, there is no compatibility between WGC and WCCC. On the contrary, two the black holes considered here can be good candidates for the  description of dark energy and dark matter, they create a good compatibility between the two above mentioned theories. Finally we want to say that, the consequence of the compatibility of the weak gravity conjecture with the weak cosmic censorship conjecture is that there are black holes in nature that repel each other.\\\\
{\bf Acknowledgments}\\
The authors would like to thank the referee for the fruitful comments and suggestions  to improve the  corresponding article.

\end{document}